\documentclass[12pt]{article}
\begin{document}
\title{{\bf BLACK HOLE CONFIGURATIONS
\\ WITH TOTAL ENTROPY LESS THAN A/4}
\thanks{Alberta-Thy-10-00, gr-qc/0101009}}
\author{
Don N. Page
\thanks{Internet address:
don@phys.ualberta.ca}
\\
CIAR Cosmology Program, Institute for Theoretical Physics\\
Department of Physics, University of Alberta\\
Edmonton, Alberta, Canada T6G 2J1
}
\date{(2000 Dec. 31)}
\maketitle
\large

\begin{abstract}
\baselineskip 16 pt

	If one surrounds a black hole
with a perfectly reflecting shell
and adiabatically squeezes the shell inward,
one can increase the black hole area $A$
to exceed four times the total entropy $S$,
which stays fixed during the process.
$A$ can be made to exceed $4S$
by a factor of order unity
before the one enters the Planck regime
where the semiclassical approximation
breaks down.
One interpretation is that the black hole
entropy resides in its thermal atmosphere,
and the shell restricts the atmosphere
so that its entropy is less than $A/4$.

\end{abstract}
\normalsize
\baselineskip 16 pt
\newpage

	The Generalized Second Law (GSL)
\cite{GSL}
of black hole thermodynamics states that
the total entropy $S$ does not decrease,
and it further states that for Einsteinian gravity
(to which this paper will be restricted,
though the generalization to various
other theories should be straightforward),
$S$ is the GSL entropy
 \begin{equation}
 S_{\rm GSL} \equiv {1\over 4}A + S_{\rm m},
 \label{eq:1}
 \end{equation}
where $A$ is the total event horizon area
of all black holes
and $S_{\rm m}$ is the entropy of matter
outside the black holes.
(I am using Planck units in which $\hbar$,
$c$, $4\pi\epsilon_0$, Boltzmann's constant $k$,
and the renormalized Newtonian gravitational constant $G$
are all set equal to unity.)
Although the Generalized Second Law
has only been proved under restricted conditions,
such as for quasistationary semiclassical
black holes
\cite{FP},
it is believed to have greater generality,
such as to rapidly evolving black holes.

	On the other hand, it is obvious that
the GSL would not apply in the form above
if the matter entropy $S_{\rm m}$ were taken to be its
von Neumann fine-grained entropy
 \begin{equation}
 S_{\rm vN} = - tr \rho \ln \rho,
 \label{eq:2}
 \end{equation}
and if information is not fundamentally lost
when a black hole forms and evaporates,
since the initial and final fine-grained entropy
of the matter could be zero when $A=0$,
whereas the GSL entropy $S_{\rm GSL} = A/4 + S_{\rm m}$,
which ignores quantum correlations or entanglements
between the black hole and the matter outside,
would be positive when (and only when)
the black hole exists and has positive area.

	Nevertheless, under some suitable
assumptions of coarse graining
(such as ignoring entanglements between
black holes and matter outside,
and ignoring complex entanglements
between different parts of the matter
emitted from a black hole if the
emission process is indeed described
by a quantum unitary process),
it has generally been assumed that it is a good
approximation to take the total entropy $S$
to be $S_{\rm GSL}$, the sum of $A/4$ and $S_{\rm m}$.

	An implicit further assumption
that is often made is that the matter entropy $S_{\rm m}$
cannot be negative (as indeed it could not,
for example, if it were given by
$S_{\rm vN} = - tr \rho \ln \rho$).
This assumption, plus the GSL,
leads to the conclusion that the total
entropy is bounded below by one-quarter
the total event horizon area:
 \begin{equation}
 {1\over 4}A \leq S.
 \label{eq:3}
 \end{equation}

	Here I shall show that the inequality
(\ref{eq:3}) can be violated.
This violation can be interpreted as either
a violation of the Generalized Second Law
(if $S_{\rm m}$ is assumed to be restricted
to nonnegative values)
or as an indication that the matter entropy
$S_{\rm m}$ must be allowed to take negative values
in order to conform to the GSL.

	Briefly, a violation of
the inequality (\ref{eq:3}) can be produced
as follows:
Take a Schwarzschild black hole
of initial mass $M_i$ and radius $2M_i \gg 1$
(in the Planck units used herein)
with negligible matter outside
(e.g., before the black hole has had time
to radiate significantly).
Assuming the GSL for this initial state,
the initial entropy $S_i$ is roughly
$A_i/4 = 4\pi M_i^2 \gg 1$,
one-quarter the initial area of the hole,
since the initial matter entropy $S_{\rm m}$
is negligible in comparison.
(If one considers as matter the thermal atmosphere
that forms when the horizon forms
in the near-horizon region $r-2M \ll 2M$,
either the entropy of this atmosphere
should be considered negligible
if it is considered to be part of $S_{\rm m}$,
or it should be considered to be part
of the black hole entropy $A/4$;
one gets too large a value for the total
entropy if one counts both $A/4$
and a large entropy associated with
the near-horizon thermal atmosphere.)

	Now surround the black hole by
a spherical perfectly reflecting shell
at a radius $r_i$ that is a few times
the Schwarzschild radius $2M_i$ of the
black hole.  This region will soon fill
up with thermal Hawking radiation
to reach an equilibrium state
of fixed energy $M_i$ inside the shell,
but for $M_i \gg 1$, all but a negligible
fraction ($\sim 1/M_i^2$ in Planck units)
of the energy will remain in the hole,
which can thus be taken still to have mass $M_i$.
Outside the shell, one will have essentially
the Boulware vacuum state with zero entropy
(plus whatever apparatus that one
will use to squeeze the shell in the next step,
but this will all be assumed to be in a pure state
with zero entropy).

	Next, squeeze the shell inward.
If this is done sufficiently slowly,
this should be an adiabatic process,
keeping the total entropy fixed.
Also, the outside itself should remain
in a zero-entropy pure state,
since the perfectly reflecting shell
isolates the region outside
from the region inside
with its black hole and thermal radiation,
except for the effects of the gravitational
field, which will be assumed to produce
negligible quantum correlations between
the inside and the outside of the shell
(as one would indeed get in a semiclassical
approximation in which the geometry
is given by a spherically symmetric classical metric).
Some of the thermal Hawking
radiation will thus be forced into the black hole,
increasing its area.

	So long as the shell is not taken
into the near-horizon region $r-2M \ll 2M$,
the radiation forced into the black hole
will have negligible energy and so will
not increase the black hole area significantly
above its initial value $A_i$.
(Indeed, some of this tiny increase in the area
just compensates for the tiny decrease
in the black hole area when it filled
the region $r < r_i$ with thermal radiation.)

	However, nothing in principle prevents one
from squeezing the shell into the near-horizon region,
where a significant amount of the near-horizon
thermal radiation can be forced into the hole,
increasing its mass $M$ and area $A = 4\pi M^2$
significantly.  Since the entropy $S$ should not change
by this adiabatic process, it remains very nearly at
$A_i/4$.  Therefore, one ends up with a squeezed
black hole configuration with $A > 4S \approx A_i$,
or total entropy significantly less than $A/4$.

	If the squeezing of the shell is accomplished
purely by tensile forces within the shell that reduce
its area, there is a minimum to the area of the shell
(where the tensile forces needed to hold the shell in place
diverge) and a maximum to the area of the squeezed
black hole, such that $A - 4S$ cannot rise above some
constant (depending on the radiation constant
and hence on the number of matter fields present)
times $M_i$ (or some related constant times $\sqrt{S}$).
However, if the squeezing is accomplished by applying
radial forces to the shell, it can be squeezed past
its minimum area to a sequence of configurations
in which both its area and the area of the black
hole inside increase by a fraction of order unity,
with now the limitation being the onset of Planck-scale
curvatures.
To borrow the language from another field,
when it is time to push, the black hole dilates.

	Perhaps the simplest way to incorporate these
$S < A/4$ configurations into black hole thermodynamics
is modify the Generalized Second Law to state that
 \begin{equation}
 \tilde{S}_{\rm GSL} \equiv S_{\rm bh} + S_{\rm m}
 \label{eq:4}
 \end{equation}
does not decrease for a suitably coarse-grained
nonnegative $S_{\rm m}$ and for a suitable definition
of $S_{\rm bh}$ that reduces to $A/4$ (in Einstein gravity)
when there are no constraints on the near-horizon
thermal atmosphere but which is less than $A/4$
when the atmosphere is constrained.
One might interpret $S_{\rm bh}$ as arising
entirely from the near-horizon thermal atmosphere,
so that if the atmosphere is unconstrained in the vertical
direction, its entropy is at least approximately $A/4$.
(There is no fundamental difficulty in allowing that
in this unconstrained case, $S_{\rm bh}$ might also have other
smaller correction terms, such as a logarithm of the number
of fields or a logarithm of $A$ or of some other black hole
parameter.  It is just that in the unconstrained case,
the leading term of $S_{\rm bh}$ should be proportional to $A$,
and the coefficient should be $1/4$, at least in Einstein gravity.)
But if the near-horizon thermal atmosphere is constrained,
it has less entropy.

	An alternative way to incorporate these
$S < A/4$ configurations is to retain the
Generalized Second Law in the original form of Eq. (\ref{eq:1}),
which is the special case of Eq. (\ref{eq:4})
in which $S_{\rm bh} = A/4$,
but now to allow $S_{\rm m}$ to become negative
when one squeezes the black hole.
For example, one might use Eq. (\ref{eq:1})
not to define $S_{\rm GSL}$ in terms of $A/4$ and $S_{\rm m}$,
but instead to define $S_{\rm m}$ as the total
entropy $S_{\rm GSL}$ minus the black hole entropy $A/4$.
(Of course, this procedure would make the GSL useless for
telling what the total entropy is, so then $S_{\rm GSL}$
would have to be found by some other procedure.)

	An analogue in which such a definition would give negative
entropies for some subsystems would be the case in which
one used the von Neumann fine-grained entropy (\ref{eq:2})
for the entropy of a total system (analogous to $S_{\rm GSL}$)
and for the entropy of one subsystem (analogous to $S_{\rm bh} = A/4$)
and then simply defined the entropy of the second subsystem
(analogous to $S_{\rm m}$) to be the total system entropy
minus the first subsystem entropy.
If the two subsystems making up the total system
have sufficient quantum correlations or entanglements,
the entropy of the second subsystem thus defined
can be negative.
For example, if the complete system is in a pure state
that entangles the two subsystems,
the von Neumann fine-grained entropy of the total system
would be zero, whereas the first subsystem would
be in a mixed state with positive von Neumann entropy,
so by the definition above, the entropy of the second
subsystem would be the negative of that positive quantity.

	In pursuing this analogue,
one could certainly take $S_{\rm GSL}$ to be the
von Neumann fine-grained entropy of the total universe,
though then under unitary evolution (e.g., no information
loss in black hole formation and evaporation),
this entropy would remain constant, so the
Generalized Second Law would be rather trivial.
(Or one could say the nontriviality is all in the
fact that the evolution is unitary.)
But the analogue would almost certainly break down for
the black hole entropy $S_{\rm bh}$ if it is
assumed to be $A/4$, since the von Neumann fine-grained entropy
of a black hole would not in general equal $A/4$.
One would expect it to be approximately $A/4$
when the black hole is maximally mixed for its area,
but, at least under the assumption of unitary
evolution, the actual von Neumann entropy
of a black hole subsystem formed from some system
of significantly smaller entropy
(e.g., from the collapse of a star)
would be expected to be much smaller than this
maximum value of approximately $A/4$
until the black hole has emitted radiation
(with which it thus becomes entangled)
with entropy at least as great as $A/4$
of the remaining black hole
\cite{PagPRL}.

	For example, suppose we take a star
of ten solar masses, or $M \sim 10^{39}$
in Planck units.  Its initial entropy will
be of the order of the number of nucleons
(of mass $m_n \sim 10^{-19}$) that it has,
$S_i \sim M/m_n \sim 10^{58}$.
If this star collapses into a black hole
of the same mass (ignoring the mass ejection
that would realistically take place),
$A/4 = 4\pi M^2 \sim 10^{79}$,
but the von Neumann entropy would remain
near $10^{58}$, 21 orders of magnitude smaller,
for a very long time,
$\sim 10^3 M S_i \sim 10^{100} \sim 10^{49} {\rm yr}$,
before the black hole increases its von Neumann entropy
significantly above $S_i$ by emitting,
and become entangled with, radiation of
significantly more entropy than $S_i$.
If one squeezed the black hole so that $A-4S \sim S$
and assumed that the original GSL Eq. (\ref{eq:1})
were valid, one would need $S_{\rm m} \sim - 10^{79}$,
which is not only negative but also is about
21 orders of magnitude in size larger than
the relevant von Neumann entropies
of the complete system and of the black hole
that are of the order of $S_i$.
Therefore, it is rather hard to interpret
such an enormous negative $S_{\rm m}$
as arising from differences of
these two von Neumann entropies,
though it still might be possible
in terms of field correlations across
a region of the height of the constrained
near-horizon thermal atmosphere.

	Let us now try to estimate
what the total entropy is of an uncharged, nonrotating
black hole configuration of mass $M$ and area $A = 16\pi M^2$,
in equilibrium with Hawking thermal radiation inside
a perfectly reflecting pure-state shell
of radius R and local mass $\mu$,
outside of which one has vacuum.
We shall take a semiclassical approximation
with a certain set of matter fields,
which for simplicity will all be assumed to be massless
free conformally coupled fields.
Given the field content of the theory,
the three parameters $(M,R,\mu)$ determine the configuration,
though the entropy should depend only on $M$ and $R$,
since the shell and the vacuum outside have zero entropy.

	It is convenient to replace the shell radius $R$
with the classically dimensionless parameter
 \begin{equation}
 W = {2M \over R},
 \label{eq:5}
 \end{equation}
which would be 0 if the shell were at infinite radius
(though before one reached this limit the black hole
inside the shell would become unstable to evaporating away)
and 1 if the shell were at the black hole horizon
(though in this limit the forces on the shell
would have to be infinite).
Then we would like to find $S(M,W)$.

	If $W$ is neither too close to 0 nor to 1,
the entropy will be dominated by $A/4 = 4\pi M^2$.
The dominant relative correction to this will come
from effects of the thermal radiation and vacuum
polarization around the hole and so would have
a factor of $\hbar$ if I were using gravitational
units $(c=G=1)$ instead of Planck units $(\hbar=c=G=1)$.
In gravitational units,
$\hbar$ is the square of the Planck mass,
so to get a dimensionless quantity from that,
one must divide by $M^2$
(or by $R^2$, which is just $4 M^2/W^2$
with $W$ being of order unity);
for the free massless fields under consideration,
there are no other mass scales in the problem
other than the Planck mass.
Therefore, in Planck units,
the first relative correction to $A/4$
will have a factor of $1/M^2$
and hence give an additive correction term
to $4\pi M^2$ that is of the zeroth power of $M$,
a function purely of $W$.
One might expect that if one proceeded
further in this way, one would find
that the entropy $S$ is given by $4\pi M^2$
times a whole power series in $1/M^2$,
with each term but the zeroth-order one
having a coefficient that is a function of $W$.
If we had been considering the possibility
of massive fields, then
these coefficients of the various powers
of $1/M^2$ would not be purely functions of $W$
but would also be functions of the masses of the fields.
However, for simplicity we shall consider only
the free massless field case here.

	In fact, I shall consider only the first
two terms in this power series,
 \begin{eqnarray}
 S(M,W) &=& 4\pi M^2 + f_1(W) + f_2(W) M^{-2} + f_3(W) M^{-4} + \cdots
 \nonumber \\
 &\approx& 4\pi M^2 + f_1(W).
 \label{eq:6}
 \end{eqnarray}
The function $f_1(W)$ will depend on the
massless matter fields present in the theory,
most predominantly through the radiation constant
 \begin{equation}
 a_r = {\pi^2 \over 30}(n_b + {7 \over 8} n_f),
 \label{eq:7}
 \end{equation}
where $n_b$ is the number of bosonic helicity states
and $n_f$ is the number of fermionic helicity states
for each momentum.
It also proves convenient to define
 \begin{equation}
 \alpha \equiv {a_r \over 384\pi^3}
  = {n_b + {7 \over 8} n_f \over 11\,520\pi},
 \label{eq:8}
 \end{equation}
which makes the entropy density of the thermal Hawking radiation
far from the hole (when $R \gg M$ or $W \ll 1$) simply $\alpha/M^3$,
and also to set
 \begin{equation}
 f_1(W) = -32\pi\alpha s(W),
 \label{eq:9}
 \end{equation}
where $s(W)$ depends (weakly) only on the ratios of the numbers
of particles of different spins and so stays fixed
if one doubles the number of each kind of species.
Then my truncated power series expression for
the entropy of an uncharged spherical black hole of
mass $M$ at the horizon (and hence horizon radius $2M$
and horizon area $4\pi M^2$) surrounded by a
perfectly reflecting shell of radius $R = 2M/W$ is
 \begin{equation}
 S(M,W) \approx 4\pi M^2 - 32\pi\alpha s(W)
  = {1\over 4} A - 32\pi\alpha s(W),
 \label{eq:9b}
 \end{equation}

	Now I shall evaluate an approximate expression
for $s(W)$ when the perturbation to the Schwarzschild
geometry is small from the thermal radiation inside the shell
and from the vacuum polarization inside and outside the shell.
There will be an additive constant to $s(W)$, giving
an additive constant to the entropy, that I shall not
be able to evaluate, but for simplicity and concreteness
I shall assume that $s(1/2)=0$,
so that the entropy is $A/4$ when the shell is at $W=1/2$ or $R=4M$.

	First, I shall ignore the Casimir energy and related effects
of the shell itself on the fields.  I would expect that these
effects would give additive corrections to $s(W)$ that are of order
$W$ or smaller (and so never large compared with unity),
whereas the leading term in the perturbative approximation
for $s(W)$ will go as $1/W^3$
(proportional to the volume inside the shell)
for $W \ll 1$ (shell radius $R \gg 2M$)
and as $1/(1-W)$ (inversely proportional to the redshift factor
to infinity) for $1-W \ll 1$ (shell radius relatively near the horizon),
so one or other of these leading terms
will dominate when $W$ is near 0 or 1.
Therefore, I shall take the stress-energy tensor inside the shell
to be approximately that of the Hartle-Hawking state
in the Schwarzschild geometry,
and that outside the shell to be approximately that
of the Boulware vacuum.

	The first part of the analysis will be done in
a coordinate system $(x^0,x^1,x^2,x^3) = (t,r,\theta,\phi)$
in which the spherically symmetric classical metric
has, at each stage of the process, the approximately static form
 \begin{equation}
 ds^2 = - e^{2\phi} dt^2 + U^{-1} dr^2
 	+ r^2 (d\theta^2 + \sin^2\theta d\varphi^2)
 \label{eq:10}
 \end{equation}
with
 \begin{equation}
 e^{2\phi} = e^{2\psi} U
 \label{eq:11}
 \end{equation}
and
 \begin{equation}
 U = 1 - {2m\over r} = 1-w
 \label{eq:12}
 \end{equation}
with
 \begin{equation}
 w \equiv {2m\over r} = 1-U = 1-(\nabla r)^2.
 \label{eq:13}
 \end{equation}
Here $\phi$, $U$, $\psi$, $m$, and $w$
are all functions of the $x^1=r$ coordinate alone,
although they also have a global dependence on
the black hole mass $M$ (the value of $r/2$
where $e^{2\phi}=0$), and (for $r>R$)
on the radius $r=R=2M/W$ of the shell and
on the total rest mass energy $\mu$ of the shell.
The Einstein equations then give
 \begin{equation}
 {d\psi\over dr} = 4\pi r (\rho + P) U^{-1}
 \label{eq:14}
 \end{equation}
and
 \begin{equation}
 {dm\over dr} = 4\pi r^2 \rho,
 \label{eq:15}
 \end{equation}
where
 \begin{equation}
 \rho = - \langle T^0_0(r)\rangle 
 \label{eq:16}
 \end{equation}
is the expectation value of the energy density
in the appropriate quantum state, and
 \begin{equation}
 P = \langle T^1_1(r)\rangle 
 \label{eq:17}
 \end{equation}
is the corresponding expectation value of the
radial pressure, both functions of $r$.
The functional form of the expectation value
of the tangential pressure
$\langle T^2_2(r)\rangle = \langle T^3_3(r)\rangle$
would then follow from the conservation
of $\langle T^{\mu}_{\nu}\rangle$ but will not be explicitly
needed in this paper.

	Since we are assuming that the
state of the quantum fields inside the shell
($r < R$) is the Hartle-Hawking
\cite{HH}
thermal state, for $r < R$ we have
 \begin{equation}
 \rho = \rho_H(M,r) \equiv {3\alpha\over 32 \pi M^4}\,\varepsilon_H(w)
 \label{eq:18}
 \end{equation}
and
 \begin{equation}
 P = P_H(M,r) \equiv {\alpha\over 32\pi M^4}\,p_H(w),
 \label{eq:19}
 \end{equation}
where on the extreme right hand side of each
of these two equations I have factored out
the dependence on the black hole mass $M$
from that on the classically dimensionless
radial function $w \equiv 2m/r$,
thereby defining two classically dimensionless
functions of $w$, $\varepsilon_H(w)$ and $p_H(w)$.

	Similarly, we are assuming that
the state of the quantum fields outside the shell
($r > R$) is the Boulware
\cite{Boul}
vacuum state, so for $r > R$ we have
 \begin{equation}
 \rho = \rho_B(M_{\infty},r)
  \equiv {3\alpha\over 32 \pi M_{\infty}^4}\varepsilon_B(w)
 \label{eq:20}
 \end{equation}
and
 \begin{equation}
 P = P_B(M_{\infty},r) \equiv {\alpha\over 32\pi M_{\infty}^4}p_B(w),
 \label{eq:21}
 \end{equation}
thereby defining two new classically dimensionless
functions of $w$, $\varepsilon_B(w)$ and $p_B(w)$.
Here
 \begin{equation}
 M_{\infty} \equiv m(r=\infty)
 \label{eq:22}
 \end{equation}
is the ADM mass at radial infinity.

	In some cases below we shall assume that there
is some extra apparatus in the region $r > R$ holding
the shell in.  If so, its energy density and radial pressure
can simply be included in $\rho_B$ and $P_B$.
In any case, we shall assume that whatever is outside
the shell is in a pure state with zero entropy.

	Below we shall also need the vacuum polarization
part of the stress-energy tensor inside the shell,
whose components I shall denote by
 \begin{eqnarray}
 &&\rho_V(M,r) \equiv \rho_H(M,r) - \rho_T(M,r)
 \nonumber \\
 &&\equiv {3\alpha\over 32 \pi M^4}\varepsilon_V(w)
 \equiv {3\alpha\over 32 \pi M^4}(\varepsilon_H(w) - \varepsilon_T(w))
 \label{eq:23}
 \end{eqnarray}
and
 \begin{eqnarray}
 &&P_V(M,r) \equiv P_H(M,r) - P_T(M,r)
 \nonumber \\
 &&\equiv {\alpha\over 32 \pi M^4}p_V(w)
 \equiv {\alpha\over 32 \pi M^4}(p_H(w) - p_T(w)),
 \label{eq:24}
 \end{eqnarray}
where $\rho_T$ and $P_T$ denote the components
of the thermal parts.

	I shall assume that the vacuum polarization part
is what the Boulware state would give if one had it
inside the shell, so that $\rho_V$ and $P_V$
have the same dependence on the local mass $m(r)$
and radius $r$ as $\rho_B$ and $P_B$ do outside
the shell (when there is no extra apparatus there).
In the first-order (in $\alpha/M^2$)
perturbative calculation being done here,
the expectation value of the stress tensor
is already first order (except possibly for that of the shell),
so its functional dependence
on $m$ can be replaced by its dependence on its
zeroth approximation, which is the black hole mass $M$ for $r < R$
and the ADM mass $M_{\infty}$ for $r > R$.
Therefore, to sufficient accuracy for our purposes,
$\rho_V$ and $P_V$ can be evaluated by using
Eqs. (\ref{eq:20}) and (\ref{eq:21}) for $\rho_B$ and $P_B$
with the ADM mass $M_{\infty}$, which is approximately
the value of the local mass $m(r)$ anywhere outside the massive shell,
replaced by the black hole mass $M$,
which is approximately the value of $m(r)$ anywhere inside the shell.
In particular, this implies that we can use
 \begin{equation}
 \varepsilon_V(w) = \varepsilon_B(w)
 \label{eq:25}
 \end{equation}
and
 \begin{equation}
 p_V(w) = p_B(w)
 \label{eq:26}
 \end{equation}

	For explicit approximate calculations,
it is useful to have explicit approximate formulas
for these various components of the stress-energy tensor,
given by the equations above from the six functions
$\varepsilon_H(w)$, $\varepsilon_B(w)$,
$\varepsilon_T(w) = \varepsilon_H(w) - \varepsilon_B(w)$,
$p_H(w)$, $p_B(w)$, and $p_T(w) = p_H(w) - p_B(w)$.
For simplicity and concreteness, I shall use those
obtained for a conformally invariant massless scalar field
in the gaussian approximation
\cite{Page82}, which gives
 \begin{eqnarray}
 \varepsilon_H(w) &\equiv& {32\pi M^4\over 3\alpha}\rho_H
 \equiv {(8\pi M)^4\over a_r}\rho_H
 \approx {1-(4-3w)^2 w^6\over (1-w)^2} - 24w^6
 \nonumber \\
 &=& 1+2w+3w^2+4w^3+5w^4+6w^5-33w^6,
 \label{eq:27}
 \end{eqnarray}
 \begin{eqnarray}
 \varepsilon_B(w) &\equiv& {32\pi M_{\infty}^4\over 3\alpha}\rho_B
 \equiv {(8\pi M_{\infty})^4\over a_r}\rho_B
 \approx {-(4-3w)^2 w^6\over (1-w)^2} - 24w^6
 \nonumber \\
 &=& -{1\over (1-w)^2}+1+2w+3w^2+4w^3+5w^4+6w^5-33w^6, 
 \label{eq:28}
 \end{eqnarray}
 \begin{equation}
 \varepsilon_T(w) = \varepsilon_H(w) - \varepsilon_B(w)
 \approx {1\over (1-w)^2} = {1\over U^2},
 \label{eq:29}
 \end{equation}
 \begin{eqnarray}
 p_H(w) &\equiv& {32\pi M^4\over \alpha}P_H
 \equiv {(8\pi M)^4\over 3 a_r}P_H
 \approx {1-(4-3w)^2 w^6\over (1-w)^2} + 24w^6
 \nonumber \\
 &=& 1+2w+3w^2+4w^3+5w^4+6w^5+15w^6,
 \label{eq:30}
 \end{eqnarray}
 \begin{eqnarray}
 p_B(w) &\equiv& {32\pi M_{\infty}^4\over \alpha} P_B
 \equiv {(8\pi M_{\infty})^4\over 3 a_r} P_B
 \approx {-(4-3w)^2 w^6\over (1-w)^2} + 24w^6
 \nonumber \\
 &=& -{1\over (1-w)^2}+1+2w+3w^2+4w^3+5w^4+6w^5+15w^6, 
 \label{eq:31}
 \end{eqnarray}
and
 \begin{equation}
 p_T(w) = p_H(w) - p_B(w)
 \approx {1\over (1-w)^2} = {1\over U^2}.
 \label{eq:32}
 \end{equation}
Note that this approximation gives
 \begin{equation}
 \rho_T \approx 3P_T \approx a_r T_{\rm local}^4,
 \label{eq:33}
 \end{equation}
just like thermal radiation in flat spacetime,
where $T_{\rm local}$ is the local value of
the Hawking temperature,
 \begin{equation}
 T_{\rm local} \approx {1 \over 8\pi m}(1-{2m\over r})^{-1/2}.
 \label{eq:34}
 \end{equation}

	Now we use the Einstein equations
(\ref{eq:14}) and (\ref{eq:15})
with the appropriate $\rho$ and $P$
on the right hand side, and with the metric function $U$
there taking on its approximate Schwarzschild form,
$1-2M/r$ for $r < R$ and $1-2M_{\infty}/r$ for $r > R$.

	We also need to consider the effect of the shell,
which has a surface stress-energy tensor with components
 \begin{equation}
 S^0_0 = - {\mu\over 4\pi R^2}
 \label{eq:35}
 \end{equation}
and
 \begin{equation}
 S^2_2 = S^3_3 = - {F\over 2\pi R},
 \label{eq:36}
 \end{equation}
where $\mu$ is the total local mass of the shell,
the shell area $4\pi R^2$ multiplied by the
local mass-energy per area $-S^0_0$ as seen by a local
observer fixed on the shell,
and $F$ is the local total tensile force pulling
together the two hemispheres of the shell,
the circumference $2\pi R$ multiplied by
the local surface tension (tensile force per length)
$-S^2_2 = -S^3_3$.

	If one integrates the Einstein equations
(\ref{eq:14}) and (\ref{eq:15}) through the shell
and uses the conservation law for the stress-energy tensor,
one get the junction conditions
\cite{junc}
in the static case that
 \begin{equation}
 \mu = R (U_{-}^{1/2} - U_{+}^{1/2})
 \label{eq:37}
 \end{equation}
and
 \begin{equation}
 8F = {\mu\over R} + (1+8\pi R^2 P_{-})U_{-}^{-1/2}
  - (1+8\pi R^2 P_{+})U_{+}^{-1/2},
 \label{eq:38}
 \end{equation}
where
 \begin{equation}
 U_{-} = 1 - {2M_{-}\over R}
 \label{eq:39}
 \end{equation}
is the value of $U$ just inside the shell ($r = R-$),
where the local mass function $m$ takes on the value $M_{-}$,
and
 \begin{equation}
 U_{+} = 1 - {2M_{+}\over R}
 \label{eq:40}
 \end{equation}
is the value of $U$ just outside the shell ($r = R+$),
where the local mass function $m$ takes on the value $M_{+}$.
Similarly, $P_{-}$ and $P_{+}$ are the expectation values of the
radial pressure of the respective quantum states just inside
and just outside the shell.

	Thus we have at least five relevant masses
for the configuration:  the black hole mass $M = m(r=2M)$,
the mass $M_{-} = m(r=R-)$ just inside the shell,
the local mass (or local energy) $\mu$
of the shell itself at radius $r=R$,
the mass $M_{+} = m(r=R+)$ just outside the shell,
and the ADM mass $M_{\infty} = m(r=\infty)$ at radial infinity.
Since the stress-energy tensor inside the shell
is that of the Hartle-Hawking state determined by $M$ and $r$,
$M_{-}$ is a function of $M$ and $R$.
Similarly, since the stress-energy tensor outside the shell
is that of the Boulware state determined by $M_{\infty}$ and $R$
(at least when we do not have an extra apparatus there
to hold the shell in place),
$M_{+}$ is a function of $M_{\infty}$ and $R$.
The junction condition (\ref{eq:27}) then gives $\mu$
as a function of $M_{-}$, $M_{+}$, and $R$,
and hence as a function of $M$, $M_{\infty}$, and $R$.
One can in principle invert this to get $M_{\infty}$
(and hence all the other masses as well)
as a function of $M$, $R$, and $\mu$,
or to get $M$ and all other masses as a function
of $M_{\infty}$, $R$, and $\mu$.
The main point is that if we just have a black hole
with the Hartle-Hawking thermal state inside a shell,
and the Boulware vacuum state outside the shell,
the semiclassical configuration (for fixed field
content of the quantum field theory)
is determined by three parameters,
though only two of them
(say $M$ and either $R$ or $W=2M/R$)
are relevant for the entropy
which resides purely inside the shell.

	To evaluate the function $s(W)$
in the truncated entropy formula (\ref{eq:9b}),
I shall consider an adiabatic process of slowly
squeezing the shell, keeping the total entropy constant
and thereby getting
 \begin{equation}
 {ds\over dW} = {M\over 4\alpha}{dM\over dW}
 \label{eq:41}
 \end{equation}
during this process.
Since this process is not strictly static,
one cannot use precisely the static metric (\ref{eq:1})
with $\phi$ and $U$ (or $\psi$ and $m$) that are
purely functions of $r$
and obey the static Einstein equations
(\ref{eq:14}) and (\ref{eq:15}).
However, one can consider a quasi-static metric
in which $\phi$ and $U$ (or $\psi$ and $m$)
have a very slow dependence on the time coordinate $t$
and the Einstein equations are only slightly different
from Eqs. (\ref{eq:14}) and (\ref{eq:15}).

	The specific calculation which I shall do
will be to have the shell squeeze itself inward by using
its own internal energy, so that no apparatus is used
outside the shell to push it inward,
and that region has only the Boulware vacuum polarization.
The contraction of the shell is assumed to be so slow
that it does not excite the vacuum outside it
but rather leaves it in the Boulware vacuum state
with constant $M_{\infty}$.
However, as the shell moves in, it is enlarging
the Boulware state region, so effectively the shell
must be creating a larger volume of vacuum
with its vacuum polarization.
This means that in the slowly inmoving frame of the shell,
there is a flux of energy from the shell into the Boulware region,
needed to enlarge the Boulware region
while keeping it static where it already exists.
[For the stress-energy tensor components of
the Boulware vacuum given by Eqs. (\ref{eq:28})
and (\ref{eq:31}), this energy influx into
the Boulware region is actually negative,
so it increases the energy of the shell as it moves inward.]

	Similarly, if the inside of the shell
were also vacuum that did not get excited by
the adiabatic contraction of the shell,
there would be a swallowing up of part of the vacuum
region by the shell as it moves inward.
This would give a flux of (negative) energy
from the vacuum inside into the shell,
decreasing its energy.
Surely this flux into the shell also
exists even if the inside is not vacuum,
and I assume that it is given by the vacuum
polarization part of the actual stress-tensor there,
which I take to be approximately that of a Boulware state
with the same $m$ and $r$.
The remaining part of the total stress-energy tensor
there, which I am calling the thermal part,
and which is given approximately by Eq. (\ref{eq:33}),
should simply be reflected by the shell and
not give an energy flux into it
(in the frame of the slowly contracting shell),
though it will contribute to the force that needs
to be counterbalanced very nearly precisely
to obey the static junction equation (\ref{eq:38})
to high accuracy in order that the shell not have any
significant acceleration relative to a static frame.

	In other words, I am assuming that if a shell
moves inward through a static geometry, the vacuum
polarization part of the stress-energy tensor will stay static,
with $T_0^0 = - \rho_V(M,r)$, $T_1^1 = P_V(M,r)$,
and $T_0^1 = T_1^0 = 0$ inside the shell, and with
$T_0^0 = - \rho_B(M_{\infty},r)$, $T_1^1 = P_B(M_{\infty},r)$,
and $T_0^1 = T_1^0 = 0$ outside the shell.
Then as the shell moves through this static stress-energy
tensor, in the frame of the shell, there will
be fluxes of energy into or out from the shell
on its two sides.  In constrast, I am assuming that
the thermal radiation part of the stress-energy tensor
will be perfectly reflected by the shell,
so that in the frame of the shell it will give
no energy fluxes into or out from the shell.

	There is a modification of this picture that occurs
when the inward motion of the shell squeezes thermal radiation
into the black hole so that its mass goes up.
While the hole mass is increasing, the vacuum polarization
inside the shell is not quite static but instead has
small $T_0^1$ and $T_1^0$ terms that, for sufficiently
slow adiabatic processes, are proportional to $\dot{M}$,
the coordinate time derivative of the black hole mass $M$.
In the present calculation, in which the shell is squeezing inself
inward by using its own internal energy,
the ADM mass $M_{\infty}$ stays fixed, and so
the vacuum stress-energy tensor outside the shell
stays static during the process, under my approximation
of neglecting Casimir-type boundary effects of the
shell itself on the quantum field.
For a sufficiently slow inward squeezing of the shell,
the $T_0^1$ and $T_1^0$ terms inside are small, but over
the correspondingly long time of the squeezing they contribute
an effect on the energy balance of the shell that is not
completely negligible when one contemplates squeezing the shell
to a final position very near the black hole horizon.
(My original neglect of these terms caused me considerable confusion
during early stages of this work and lectures I gave about it.)

	My present procedure for calculating the small
$T_0^1$ and $T_1^0$ terms inside the shell is to assume that
the shell squeezing, and all consequent processes,
occur so slowly that $T_0^0$ and $T_1^1$
are given to high accuracy by the same functions
of $M$ and $r$ as they are when the geometry is static,
namely $-\rho_V(M,r)$ and $P_V(M,r)$.
Then I assume that the vacuum polarization part of
the stress-energy tensor is itself conserved
away from the shell,
so one can use the conservation of its energy
to deduce the radial derivative of $e^{\psi}r^2 T_0^1$.

	In particular, if we let the vacuum polarization
part of the stress-energy tensor have the component
 \begin{equation}
 T_0^1 = {\alpha\dot{M}w^2\over 4\pi M^4}e^{-\psi}f,
 \label{eq:42a}
 \end{equation}
with the factors chosen so that $f$ is a function purely of $w$,
then $T^{\mu}_{0;\mu}=0$ becomes
 \begin{equation}
 {\partial f\over\partial r} = {\pi M^2 e^{\psi}r^2\over\alpha\dot{M}}
 [\dot{\rho}_V + {\dot{m}\over rV}(\rho_V + P_V)].
 \label{eq:42b}
 \end{equation}
For the region inside the shell with $r$ not too much larger than $2M$,
one has $m \approx M$ and $e^{\psi} \approx 1$
(possibly after suitably normalizing the time coordinate $t$).
Then if one uses Eqs. (\ref{eq:23})-(\ref{eq:26}),
one can rewrite Eq. (\ref{eq:42b}) as
 \begin{equation}
 {df\over dr} =
  -{3w\over 4}{d\over dw}\left({\varepsilon_B \over w^4}\right)
  -{3\varepsilon_B + p_B \over 8w^3(1-w)}.
 \label{eq:42c}
 \end{equation}
Given the functions $\varepsilon_B(w)$ and $p_B(w)$,
e.g., as given by Eqs. (\ref{eq:28}) and (\ref{eq:31})
from the gaussian approximation for a conformally
invariant massless scalar field, one can integrate Eq. (\ref{eq:42c})
to obtain $f(w)$ up to a constant of integration.
Although the constant of integration is not important,
it can also be fixed
by assuming that an observer that remains at fixed $w = 2m/r$
as $m$ changes sees in its frame no energy flux
in the limit that $w$ is taken to unity,
which implies that the flux of vacuum polarization energy
through the horizon is taken to be zero.

	After one calculates the vacuum polarization part
of the stress-energy tensor, which gives
$T_1^1-T_0^0 = \rho_B + P_B$ and $T_0^1 = 0$
outside the shell and
$T_1^1-T_0^0 = \rho_V + P_V$ and $T_0^1$
as given by Eq. (\ref{eq:42a}) inside the shell,
one can then calculate the fluxes of energy out from and
into the shell and insert these into the conservation
equations for the surface stress-energy tensor of the shell.
For a very slowly expanding or contracting shell, one finds that
 \begin{equation}
 d\mu = 4F dR
  + 4\pi R^2 dR[(\rho_B+P_B)U_{+}^{-1/2} - (\rho_V+P_V)U_{-}^{-1/2}]
  - 4\pi R^2 T_0^1 U_{-}^{-1/2} dt.
 \label{eq:42d}
 \end{equation}
The first term on the right hand side is the work done by
the tensile force within the shell,
and the remaining terms are the energy input from the
vacuum stress-tensor components $\rho_B$ and $P_B$
just outside the shell and the
vacuum stress-tensor components $\rho_V$, $P_V$, and $T_0^1$
just inside the shell.

	One now combines this local energy
conservation equation for the shell with the static junction
equations (\ref{eq:37}) and (\ref{eq:38})
that should still apply to high accuracy
in this slowly evolving situation to
keep the shell radius from accelerating
too rapidly.
When one also combines this with the
integrals of Eq. (\ref{eq:15}),
 \begin{equation}
 M_{-} = M + \int_{2M}^R 4\pi r^2 dr \rho_H,
 \label{eq:43}
 \end{equation}
 \begin{equation}
 M_{+} = M_{\infty} - \int_R^{\infty} 4\pi r^2 dr \rho_B,
 \label{eq:44}
 \end{equation}
one finds
 \begin{equation}
 \left(1 - {4\alpha f \over M^2}\right) dM
 \approx - 4\pi R^2 dR (\rho_T + P_T)
 \label{eq:45}
 \end{equation}
during the adiabatic contraction of the shell,
which, up to the small correction factor involving $f$,
is precisely what one would get
in flat spacetime from adiabatically
contracting a ball of thermal radiation.

	Next, we can use the fact that
$R = 2M/W$ to derive that
 \begin{equation}
 {dW \over dM} = {2\over R}\left(1-{M\over R}{dR\over dM}\right)
 \approx {2\over R}
 \left[1+{M(1-4\alpha f/M^2)\over 4\pi R^3(\rho_T + P_T)}\right],
 \label{eq:46}
 \end{equation}
where $f$ and $\rho_T + P_T$ are to be evaluated at $r=R$
or $w\approx W$.
Inserting this back into Eq. (\ref{eq:41})
then gives
 \begin{equation}
 {ds\over dW}
 \approx {3\varepsilon_B + p_B \over 4W^4}
  \left\{1+{4\alpha\over M^2}
   \left[{3\varepsilon_B + p_B\over 4W^3}-f\right]\right\}^{-1}.
 \label{eq:47}
 \end{equation}
For massless particles of any spin,
it should be a good approximation to take
$\rho_T + P_T \approx (4/3) a_r T_{\rm local}^4$ in terms of
the local temperature $T_{\rm local}$, and this implies that
$3\varepsilon_B + p_B \approx 4/(1-W)^2$, so
 \begin{equation}
 {ds\over dW}
 \approx {1 \over W^4 (1-W)^2}
  \left\{1+{4\alpha\over M^2}
   \left[{1 \over W^3 (1-W)^2}-f\right]\right\}^{-1}.
 \label{eq:47b}
 \end{equation}
 
	If we omitted the $f$ term from the radial
flux of vacuum polarization energy when $M$ changes,
as I indeed first erroneously did,
then the factor inside the curly brackets above would
diverge as one approached the horizon, where $W=1$.
This implies that the reciprocal of this factor would cancel
the divergence in the factor before it, so $ds/dW$
would stay finite all the way down to $W=1$,
and one would find that the increase of one-quarter the area over
the entropy would be limited to an amount of the order
of $\alpha M$.  For large $M$ this is large in absolute
units, but it is always much smaller than the entropy
itself, which is of the order of $4\pi M^2$.

	However, one can use the fact that
the regularity of the Hartle-Hawking stress-energy
tensor at the horizon implies that $\rho_H + P_H$,
and hence $3\varepsilon_H + p_H$, must go to zero
at least as fast as $1-w$ as one approaches the horizon.
(This is easiest to see in the Euclidean section
with imaginary time $t$,
on which for fixed coordinates $\theta$ and $\varphi$,
the horizon is at the center of a regular rotationally
symmetric two-surface with angular coordinate
$i\kappa t$ with $\kappa \approx 1/(4M)$ being the black hole
surface gravity and with the radial distance being
roughly $4M\sqrt{1-w}$ when $1-w \ll 1$.
Then $P_H = T_1^1$ is the pressure in the radial direction,
and $-\rho_H = T_0^0$ is the Euclidean pressure in the
Euclidean angular direction, and regularity at the origin
demands that the difference go to zero at least as fast
as the square of the radial distance from the origin.)
Then one can show that $f$ cancels the divergence
in $W^{-3}(1-W)^{-2}$ so that $W^{-3}(1-W)^{-2}-f$
stays finite as one approaches the horizon.
In fact, if one chooses the constant of integration of $f$
so that the flux of vacuum polarization energy through
the horizon is zero as $M$ is slowly changed,
then $W^{-3}(1-W)^{-2}-f$ actually goes to zero linearly
with $1-W$ as one approaches the horizon.
For example, using this constant of integration and
the gaussian approximation for $3\varepsilon_B$ and $p_B$
leads to
 \begin{equation}
 {ds\over dW}
 \approx {1 \over W^4 (1-W)^2}
  \left\{1+{4\alpha\over M^2 W^3}
   (1-W)(1+3W+6W^2+2W^3+7W^4+13W^5)\right\}^{-1}.
 \label{eq:47c}
 \end{equation}
 
 	Therefore, we see that the correction term
that is first order in $\alpha/M^2$ inside the curly brackets
of Eqs. (\ref{eq:47})-(\ref{eq:47c}) does not diverge
as one takes $1-W$ to zero but instead always remains small.
Therefore, we can drop it (as we have also neglected other
finite corrections that are linear in $\alpha/M^2$) and
integrate the zeroth-order part of Eq. (\ref{eq:47c})
to get an explicit formula for $s(W)$:
 \begin{equation}
 s(W) \approx \int_{1/2}^W {dw \over w^4(1-w)^2}
 = {1\over 1-W}+4\ln{W\over 1-W}-{1\over 3W^3}
 -{1\over W^2}-{3\over W}+{32\over 3}.
 \label{eq:48}
 \end{equation}
As discussed above, I arbitrarily chose the constant of integration
of this integral to make $s(W) = 0$ at $W = 1/2$ or $R = 4M$,
but this is not likely to be valid,
and there are also Casimir energy effects from the
shell and corrections to Eqs. (\ref{eq:29}) and (\ref{eq:32})
that would give correction terms at least of order $W$
and likely also of the order of a constant and of order $1/W$.

	Finally, we can insert this form for $s(W)$
into Eq. (\ref{eq:9b}) to get
 \begin{eqnarray}
 S(M,W) &=& {1\over 4} A [1 - {8\alpha\over M^2} s(W)]
 = 4\pi M^2 -32\pi\alpha s(W)
 \nonumber \\
 &\approx& 4\pi M^2 - 32\pi\alpha[{1\over 1-W}+4\ln{W\over 1-W}
        - {1\over 3W^3} - {1\over W^2} + O({1\over W})]
 \nonumber \\
 &\approx& 4\pi M^2 - 32\pi\alpha[{R\over R-2M}+4\ln{2M\over R-2M}
        - {R^3\over 24M^3} - {R^2\over 4M^2}],
 \label{eq:49}
 \end{eqnarray}
where after the last approximate equality
I have dropped the terms in Eq. (\ref{eq:48})
that I suspect are always dominated by corrections
to my approximations that I have not included.
Although I have retained four terms from $s(W)$,
only the first two terms should be kept when
$1-W = 1-2M/R \ll 1$ (shell very near the horizon),
and only the next two terms should be retained
when $W = 2M/R \ll 1$ (shell very large compared with the black hole).

	The result indicated by Eq. (\ref{eq:48})
is precisely the same that one would obtain by taking
the geometry to be Schwarzschild with a thermal bath
of radiation with local Hawking temperature
 \begin{equation}
 T_{\rm local} = {1 \over 8\pi M}(1-{2M\over r})^{-1/2}.
 \label{eq:50}
 \end{equation}
and entropy density $(4/3)a_r T_{\rm local}^3$,
and then taking the total entropy to be $4\pi M^2$
plus the entropy difference between that inside the
shell at radius $R$ and that inside the radius $4M$.
If one na\"{\i}vely integrates this assumed entropy density
all the way down to the horizon, one would get a divergence,
but one can take the attitude that this divergence is
regulated so that the entropy in this thermal atmosphere
below some radius like $4M$ (the precise value of which
doesn't matter much, since the assumed entropy density
is this region is of the order of $\alpha/M^3$)
is the black hole entropy $S_{\rm bh} \approx A/4 = 4\pi M^2$.
Then one can say that if the shell
is put at a much larger radius,
the entropy of the thermal Hawking radiation
outside $4M$ or so would be matter entropy $S_{\rm m}$
that would add to $S_{\rm bh}$,
which is certainly an uncontroversial assumption.

	What I have found from my consideration
of having the shell squeezed in adiabatically
is that if the shell is put much nearer the horizon
than a radius of $4M$ or so is, then the entropy
is correspondingly less than the usual black hole entropy
$S_{\rm bh} \approx A/4 = 4\pi M^2$.
Because the thermal atmosphere is restricted
from filling up the region to $4M$ or so,
it does not have the entropy needed to make
the total entropy as large as $A/4$.

	The next question is the range of $W$
over which one would expect that Eq. (\ref{eq:49})
is approximately valid.
For very small $W$ or very large $R$, one essentially
has a black hole of mass $M$ surrounded by
a much bigger volume, $V \sim 4\pi R^3/3$,
of radiation in nearly flat
spacetime with Hawking temperature $1/(8\pi M)^{-1}$,
energy density roughly $3\alpha/(32\pi M^4)$,
and entropy density roughly $\alpha/M^3$.
The dominant term for
the total energy of the radiation is
$E_r \sim \alpha R^3/(8M^4)$,
and from Eq. (\ref{eq:49}), the dominant term for
the total entropy of the radiation is
$S_r \sim 4\pi\alpha R^3/(3M^3)$.
This agrees with the standard expression
for the entropy of thermal radiation of energy $E_r$
in a volume $V$,
 \begin{equation}
 S_r = {4\over 3}(a_r V)^{1/4} E_r^{3/4}
 \sim {4\pi\over 3}\alpha^{1/4}(8 R E_r)^{3/4}.
 \label{eq:59}
 \end{equation}

	For fixed total energy $M_{\infty} = M + E_r \ll R$,
the total entropy
 \begin{equation}
 S \approx 4\pi M^2 + S_r
 \sim 4\pi(M_{\infty}-E_r)^2
  + {4\pi\over 3}\alpha^{1/4}(8 R E_r)^{3/4}
 \label{eq:60}
 \end{equation}
is indeed extremized for
 \begin{equation}
 E_r \sim {\alpha R^3 \over 8 M^4}
  = {\alpha R^3 \over 8(M_{\infty}-E_r)^4},
 \label{eq:61}
 \end{equation}
but the extremum is a local entropy maximum
if and only if $5E_r \leq M_{\infty}$ or $4E_r \leq M$
\cite{BHbox},
which implies that one needs $R \leq (2M^5/\alpha)^{1/3}$ or
 \begin{equation}
 W \geq \left({4\alpha\over M^2}\right)^{1/3}
 \label{eq:62}
 \end{equation}
for thermodynamic stability.

	For smaller values of $W$ (larger values of $R$),
the radiation energy $E_r$ is more than $20\%$ of the
total available energy $M_{\infty}$ (assumed to be held fixed),
and then if the black hole emits some extra radiation
and shrinks, it heats up more than the radiation does,
leading to an instability in which the black hole
radiates away completely. On the other hand,
if the black hole absorbs some extra radiation,
it will grow and cool down more than the surrounding radiation,
therefore cooling down more and absorbing more radiation,
until the radiation energy drops to the lower
positive root of Eq. (\ref{eq:61}),
which is less than $0.2 M_{\infty}$ and hence is
at least locally stable.

	At the opposite extreme,
the question is how small $1-W$ can be.
Here the fundamental limit is the Planck regime,
which is the boundary of the semiclassical
approximation being used in this paper.
The Boulware vacuum energy density $\rho_B$
just outside a massless shell
(so that the mass just outside, $M_{+}$,
is very nearly the same as the black hole mass $M$;
for positive shell mass $\mu$, $\rho_B$
would have an even greater magnitude)
is, for very small $U = 1-W$,
$\rho_B \sim - 3\alpha/(32\pi M^4 U^2)$.
Suppose the semiclassical theory is valid
until the orthonormal Einstein tensor component
$G_0^0 = -8\pi\rho_B \sim 3\alpha/(4 M^4 U^2)$
reaches a maximum value of, say, $C_M$,
which would be expected to be of order unity
(orthonormal curvature component of the order
of the Planck value).
This gives the restriction
 \begin{equation}
 U = 1-W \geq \left({3\alpha\over 4 C_M M^4}\right)^{1/2}.
 \label{eq:63}
 \end{equation}
 
	For $U = 1-W \ll 1$, the spatial distance
from the shell to the horizon is $D \sim 4M U^{1/2}$,
so this restriction on $U$ gives a minimum distance the
shell can be from the horizon:
 \begin{equation}
 D \geq \left({192\alpha\over C_M}\right)^{1/4},
 \label{eq:64}
 \end{equation}
in Planck units, as all quantities are in this paper
unless otherwise specified.

	If we combine the restriction (\ref{eq:63})
with the lower bound on $W$ from Eq. (\ref{eq:62})
and re-express the combined restriction as a restriction on
the radius $R$ of the shell, we get
 \begin{equation}
 2M+{1\over M}\sqrt{3\alpha\over C_M} \leq R
  \leq \left({2M^5\over\alpha}\right)^{1/3}.
 \label{eq:65}
 \end{equation}
Alternatively, in terms of the distance $D$
of the shell to the horizon (which is $D \sim R$
for $R \gg 2M$), we get
 \begin{equation}
 \left({192\alpha\over C_M}\right)^{1/4} \leq D
  \leq \left({2M^5\over\alpha}\right)^{1/3}.
 \label{eq:64b}
 \end{equation}

	If we now insert the restriction (\ref{eq:63})
or (\ref{eq:64}) into the asymptotic form of the total
entropy (\ref{eq:49}) for $U = 1-W \ll 1$, which is
 \begin{eqnarray}
 S(M,W) &\sim& 4\pi M^2 - {32\pi\alpha\over U}
 	\sim 4\pi M^2 \left( 1 - {128\alpha\over D^2} \right)
 \nonumber \\
	&\sim& 4\pi M_{\infty}^2 - {8\pi\alpha\over U}
	\sim 4\pi M_{\infty}^2 \left( 1 - {32\alpha\over D^2} \right),
 \label{eq:65b}
 \end{eqnarray}
we get the limitation
 \begin{equation}
 S(M,W) \geq 4\pi M^2 \left( 1 - 16 \sqrt{\alpha C_M \over 3} \right)
 = 4\pi M^2
  \left( 1 - \sqrt{C_M \over 135\pi}(n_b + {7\over 8}n_f) \right).
 \label{eq:66}
 \end{equation}
This can be re-expressed as a limitation on how much
the area $A$ of a black hole can exceed
four time the entropy, $4S$:
 \begin{equation}
 A - 4S \leq A\sqrt{C_M \over 135\pi}(n_b + {7\over 8}n_f).
 \label{eq:66b}
 \end{equation}

	Therefore, unless we have
$N \equiv n_b + 7n_f/8$, the effective number of
one-helicity particles, comparable to or greater than
$135\pi/C_M \approx 424/C_M$,
the fractional increase of the black hole area $A$
above $4S$ is restricted to be rather small,
though even just $N=4$ from two-helicity
gravitons and photons would give a fractional increase of
about 10\% if the curvature limitation $C_M$
is one in Planck units.

	However, this does raise the
interesting theoretical question of what would
happen in a theory in which the effective number $N$
of particles is so large that $N C_M/(135\pi)$
is bigger than unity.
Na\"{\i}vely it would then appear that,
without running into excessive curvatures
(i.e., $G^0_0 > C_M$),
one could put the shell sufficiently close
to the black hole that the total entropy,
given approximately by Eq. (\ref{eq:65}),
would be negative, which is surely nonsense.
Of course, one could never get to such
a configuration by adiabatically compressing
the shell, since one would then have started out
with positive entropy that would not decrease
(though it does appear that one could
in principle push hard and far enough on
the shell that the black hole could dilate
to an arbitrarily large radius for the fixed initial
entropy, which by itself would be rather remarkable).
However, one could imagine constructing
such a shell in place and then evacuating
the thermal radiation from above it,
which would seem to leave behind a black hole
configuration of negative entropy.
In our world this possibility might be
excluded by the limited number of particle
species that exist and contribute to
the thermal Hawking radiation,
but one would like to see a direct
argument of why negative total entropies
of black hole configurations cannot
be achieved even in a model
universe in which one had a huge number of fields.

	The first guess that came to me
off the top of my head is that of course
no physical shell can be perfectly reflecting.
A partially reflecting shell should be able to
squeeze enough of the Hawking radiation
into a black hole to raise its area
above four times the entropy, but
the transmission will put a limit on
how effective this process can be.
If there are more species of particles
that can be partially transmitted,
the shell may become less effective
in increasing the black hole area,
and conceivably this could offset
the increase in the otherwise theoretically
allowed fractional area increase from
the increase of the number of species.
However, this is just a wild guess,
and so the problem will be left as
an exercise for the reader.

	One refinement of the results above
that should be explained is that although
I have used a Schwarzschildean coordinate
system (\ref{eq:10}) in the analysis
described so far, this system breaks down
when one follows the Boulware vacuum sufficiently
far inward.
In particular, the Schwarzschildean radial coordinate
$r$, which is $1/(2\pi)$ times the circumference
of the two-sphere, is only a good coordinate
when it decreases monotonically inward.
This indeed occurs when $U = 1 - 2m/r = (dr/dD)^2$
remains positive, where $D$ is the proper radial distance.
But because the Boulware vacuum energy density
$\rho_B$ is negative for sufficiently small $U$,
as one moves inward with $r$ initially decreasing,
the mass function $m(r)$ increases rather than decreases,
and eventually one reaches a radius where $2m = r$
and hence $U = 0$.
In this case, $-g_{00} = e^{2\phi}$ remains positive
(which implies that $e^{2\psi} = e^{2\phi} U^{-1}$
diverges, so $\psi$ is no longer a good metric function either),
which means that one has not reached the horizon,
but rather just a location where the gradient of
the Schwarzschildean radial coordinate vanishes.
Inside this point, the gradient of $r$ reverses
sign, so now $r$ increases as one moves inward.
The Einstein equation (\ref{eq:15}) implies,
since the energy density remains negative,
that the mass function $m$ now decreases as
one moves inward.

	Since at small radii the Boulware
vacuum polarization gives a stress-energy tensor
that is dominated by contributions that look like
thermal radiation but with the opposite sign
(e.g., an isotropic pressure that is one third
the energy density, both of which are negative),
one can incorporate an approximation for its
back reaction on the metric simply by
solving the Tolman-Oppenheimer-Volkoff
equations for such a thermal fluid with
 \begin{equation}
 P_B \approx {1\over 3}\rho_B \approx -P_T
  \approx -{1\over 3}a_r T_{\rm local}^4
  \approx -{\alpha \over 32\pi M_{\infty}^4}e^{-4\phi}.
 \label{eq:67a}
 \end{equation}
If one defines the function
 \begin{equation}
 W \equiv 1 + 8\pi r^2 P_B
 \label{eq:67}
 \end{equation}
(not to be confused with the previous use
of $W$ as $2M/R$), then one gets the two differential
equations
 \begin{eqnarray}
 U(4-U-3W)dW &=& 2(W-2U)(1-W)dU
 \nonumber \\
 &=& 2(W-2U)(1-W)(4-U-3W)dr/r.
 \label{eq:68}
 \end{eqnarray}

	One can match with the Schwarzschild metric,
slightly perturbed by the Boulware vacuum polarization,
at $\sqrt{\alpha}/M_{\infty} \ll U \ll 1$,
where $1-W \approx \alpha M_{\infty}^{-2} (1-U)^{-2} U^{-2}$
and $r \approx 2M_{\infty}(1-U)^{-1}$.
Then as one integrates Eq. (\ref{eq:68}) inward,
initially $r$, $U$, and $W$ all decrease,
until $U$ and $W$ simultaneously go to zero.
This is a singular point of Eq. (\ref{eq:68}),
but one can easily see that $U$ goes to zero
quadratically with $W$, so that as $W$
crosses zero and becomes negative,
$U$ becomes positive again.
If we define the new variable
 \begin{equation}
 Y \equiv {U\over W} \equiv {1-2m/r \over 1 + 8\pi r^2 P},
 \label{eq:69}
 \end{equation}
we find that it decreases monotonically
as we go inward, starting at the small positive value
$Y = Y_0 \ll 1$ in order that one be in the regime where
Eq. (\ref{eq:67a}) is valid,
but then with $Y$ going negative where $U$ goes to zero
quadratically with $W$.

	In terms of $Y$ and $W$, the first differential
equation of (\ref{eq:68}) becomes
 \begin{equation}
 {dW\over dY} = {2(1-2Y)W(1-W)\over Y[(1+2Y)(2-W)-3YW)]}.
 \label{eq:70}
 \end{equation}
Although this equation is also singular at $Y=0$ and $W=0$,
$YdW/WdY = 1$ there, so $W$ just passes through zero
linearly with $Y$, with a positive coefficient that one can
calculate is roughly $M_{\infty}/\sqrt{\alpha}$.
Then one can easily see that the right hand side of
Eq. (\ref{eq:70}) is always positive,
since $1-W > 0$ and $1-2Y > 0$ everywhere where
Eq. (\ref{eq:67a}) is valid,
so $Y$ and $W$ are both monotonically decreasing
variables as one moves inward through the negative
pressure (and negative energy-density) vacuum
Boulware stress-energy tensor.

	Eq. (\ref{eq:70}) is not separable,
but by pulling out the separable parts, one can get
 \begin{equation}
 {W \over Y \sqrt{1-W}} \propto \exp
 \left[-\int{(8-7W)dY \over (2-W)+(4-5W)Y}\right].
 \label{eq:71}
 \end{equation}
For the initial conditions above with
$\sqrt{\alpha}/M_{\infty} \ll 1$
one can now integrate Eq. (\ref{eq:71}) approximately
in the various regimes for $W$ and match them:

	Initially, $W$ is near 1, and so
 \begin{equation}
 {W \over Y \sqrt{1-W}} \approx {M_{\infty}\over\sqrt{\alpha}}(1-Y).
 \label{eq:72}
 \end{equation}
In particular, the right hand side is correct to first order in $Y$
when $\sqrt{\alpha}/M_{\infty} \ll Y$.
Then as $W$ goes from near 1 to much less than $-1$,
$Y$ is so small and changes so little that the
integral in Eq. (\ref{eq:71}) has only a negligible
contribution.  Therefore, Eq. (\ref{eq:72}) remains approximately
valid while $W$ drops from being near 1 until it becomes sufficiently
negative that $Y$ no longer has a small magnitude,
though the right hand side is no longer correct to first order
in the small quantity $Y$; an expression correct to first order
in $Y$ for
$-\sqrt{\alpha}/M_{\infty} \ll Y \ll \sqrt{\alpha}/M_{\infty}$,
which implies $|W| \ll 1$, is
 \begin{equation}
 {W \over Y \sqrt{1-W}}
   \approx {M_{\infty}\over\sqrt{\alpha}}(1+2Y)^{-2}.
 \label{eq:73}
 \end{equation}
Finally, when $W$ is very large and negative,
$Y$ can grow from its tiny value and gives,
for $0.2 < Y \ll -\sqrt{\alpha}/M_{\infty}$,
 \begin{equation}
 {W \over Y \sqrt{1-W}}
   \approx {M_{\infty}\over\sqrt{\alpha}}(1+5Y)^{-7/5}.
 \label{eq:74}
 \end{equation}
If one only needs an expression for $W/(Y\sqrt{1-W})$
that is approximately correct but not necessarily correct
to first order in $Y$, then Eq. (\ref{eq:74}) can
be taken to apply over the whole allowed range
where Eq. (\ref{eq:67a}) is valid, namely
$-0.2 < Y \leq Y_0 \ll 1$.  One can then algebraically
solve Eq. (\ref{eq:74}) for $W \equiv 1 + 8\pi r^2 P_B$ to get
the explicit approximate formulas for $W \equiv 1 + 8\pi r^2 P_B$
and $U \equiv 1-2m/r \equiv WY$ in terms of $Y \equiv U/W$:
 \begin{equation}
 W \approx {M_{\infty}\over\sqrt{\alpha}}Y(1+5Y)^{-7/5}
   \left[1+{M_{\infty}^2\over 4\alpha}Y^2(1+5Y)^{-14/5}\right]^{1/2}
   -{M_{\infty}^2\over 2\alpha}Y^2(1+5Y)^{-14/5},
 \label{eq:74b}
 \end{equation}
 \begin{equation}
 U \approx {M_{\infty}\over\sqrt{\alpha}}Y^2(1+5Y)^{-7/5}
   \left[1+{M_{\infty}^2\over 4\alpha}Y^2(1+5Y)^{-14/5}\right]^{1/2}
   -{M_{\infty}^2\over 2\alpha}Y^3(1+5Y)^{-14/5}.
 \label{eq:75}
 \end{equation}

	One can go on to use Eqs. (\ref{eq:67a})-(\ref{eq:68})
to solve for the behavior of $r$, $m$, and $\phi$.
To avoid expressions that are too cumbersome,
it is useful to divide up the region where Eq. (\ref{eq:67a}) is valid
into three overlapping regions and give explicit approximate results
for each region separately.

	In Region 1, where $1-W \ll 1$ or
$Y \gg \sqrt{\alpha}/M_{\infty}$, one gets
 \begin{equation}
 r \approx {2M_{\infty}\over 1-Y},
 \label{eq:76}
 \end{equation}
 \begin{equation}
 m \approx M_{\infty} + {3\alpha\over M_{\infty}Y},
 \label{eq:77}
 \end{equation}
 \begin{equation}
 \phi \approx {1\over 2} \ln{Y}.
 \label{eq:78}
 \end{equation}

	In Region 2, where $U \ll 1$, which is always true
for positive $Y \leq Y_0 \ll 1$ and is true also for
negative $Y$ if $-Y \ll (M_{\infty}^2/\alpha)^{1/3}$,
one gets
 \begin{equation}
 r \approx 2M_{\infty}(1+\sqrt{4\alpha/M_{\infty}^2 + Y^2}),
 \label{eq:79}
 \end{equation}
 \begin{equation}
 m \approx M_{\infty}\left[
  1+{M_{\infty}\over 2\sqrt{\alpha}}Y^2
   +\left(1-{M_{\infty}\over 4\sqrt{\alpha}}Y
           -{M_{\infty}^2\over 4\alpha}Y^2\right)
	     \sqrt{{4\alpha\over M_{\infty}^2}+Y^2}\right],
 \label{eq:80}
 \end{equation}
 \begin{equation}
 \phi \approx {1\over 4}\ln{\alpha\over M_{\infty}^2}
  -{1\over 2} \sinh^{-1}{{M_{\infty}\over 2\sqrt{\alpha}}Y}.
 \label{eq:81}
 \end{equation}
One can see that $r$ reaches a minimum value of roughly
$r_m \approx 2M_{\infty}+4\sqrt{\alpha}$ at $Y=0$,
where the mass function $m$ attains its maximum value
$m_m = r_m/2 \approx M_{\infty}+2\sqrt{\alpha}$.
Then as one moves further inward,
to negative values of $Y$ and $W$, $r$ increases again,
and $m$ decreases.

	In Region 3, where $-W \gg 1$
or $-Y \gg \sqrt{\alpha}/M_{\infty}$, one gets
 \begin{equation}
 r \approx 2M_{\infty}(1+5Y)^{-1/5},
 \label{eq:82}
 \end{equation}
 \begin{equation}
 m \approx M_{\infty}
  - {1\over\alpha}\left({-M_{\infty}Y\over 1+5Y}\right)^3,
 \label{eq:83}
 \end{equation}
 \begin{equation}
 \phi \approx
  -{1\over 2} \ln{\left({-M_{\infty}Y\over\alpha}\right)}
  +{3\over 5} \ln{(1+5Y)}.
 \label{eq:84}
 \end{equation}
One can see that the mass function $m$ crosses zero
at $Y \approx (\alpha/M_{\infty}^2)^{1/3}$
and thereafter becomes negative, approaching negative
infinity as $Y$ approaches its lower limit of $-0.2$.

	If we solve Eq. (\ref{eq:82}) for $Y$ in terms of $r$,
then we can write the approximate metric in Region 3
in terms of explicit functions of $r$:
 \begin{eqnarray}
 ds^2 \approx &-& {5\alpha\over M_{\infty}^2}
  \left({2M_{\infty}\over r}\right)^6
  \left[1-\left({2M_{\infty}\over r}\right)^5\right]^{-1}dt^2
 \nonumber \\
 &+&{125\alpha\over M_{\infty}^2}
  \left({2M_{\infty}\over r}\right)^{14}
  \left[1-\left({2M_{\infty}\over r}\right)^5\right]^{-3}dr^2
 \nonumber \\
 &+& r^2 (d\theta^2 + \sin^2\theta d\varphi^2).
 \label{eq:85}
 \end{eqnarray}
One can see that there is a naked singularity, where
$r$ goes to $+\infty$ and $-g_{00}$ goes to zero,
at a finite proper radial distance.  In fact, for $r \gg M_{\infty}$,
the proper distance to the singularity along a geodesic with
constant $t$, $\theta$, and $\varphi$ is,
if one were to continue using this metric in a region
where the curvature it indicates is comparable to the Planck values,
 \begin{equation}
 \ell \approx \sqrt{125\alpha\over 3}
  \left({2M_{\infty}\over r}\right)^6.
 \label{eq:86}
 \end{equation}
Therefore, for $\ell \ll \sqrt{\alpha}$,
the metric (\ref{eq:85}) has the approximate form
 \begin{equation}
 ds^2 \approx -\sqrt{9\alpha\over 5}\ell{dt^2\over M_{\infty}^2}
 + d\ell^2
 + \left({125\over 9\alpha}\right)^{1/6}
   (2M_{\infty})^2 \ell^{-1/3} (d\theta^2 + \sin^2\theta d\varphi^2).
 \label{eq:87}
 \end{equation}

	A slightly cruder form for the approximate metric
of the Boulware vacuum region with backreaction,
but one which applies over the whole spacetime, uses
(in this section only; elsewhere $R$ is the circumference
of the shell divided by $2\pi$)
 \begin{equation}
 R \equiv {2M_{\infty}\over 1-e^{2\phi}}
   \equiv {2M_{\infty}\over 1+g_{00}}
 \label{eq:88}
 \end{equation}
as the independent radial variable,
since this variable, like $e^{2\phi} \equiv -g_{00}$,
varies monotonically with radial distance,
from $R = 2M_{\infty}$ at the naked singularity to
$R = \infty$ at radial infinity.  Then the metric takes
the approximate form
 \begin{eqnarray}
 ds^2 \approx &-& \left(1-{2M_{\infty}\over R}\right) dt^2
 \nonumber \\
 &+&
  \left(1+{8\alpha\over M_{\infty}^2}-{2M_{\infty}\over R}\right)^{-1}
   dR^2
 \nonumber \\
 &+& R^2
  \left[1+{6\alpha\over M_{\infty}^2(1-2M_{\infty}/R)}\right]^{1/3}
   (d\theta^2 + \sin^2\theta d\varphi^2).
 \label{eq:89}
 \end{eqnarray}

	For $\alpha/M_{\infty}^2 \ll 1$ as we have always been assuming,
this metric reduces to very nearly the Schwarzschild metric
for $R-2M_{\infty} \gg \alpha/M_{\infty}$, which includes
the region with $Y \gg Y_0$ for which Eq. (\ref{eq:67a}) is not valid,
and it also gives a reasonably good approximation to the metric
in Regions 1 and 2 where Eq. (\ref{eq:67a}) is likely to be valid.
In the part of Region 3 where $r \gg M_{\infty}$,
the metric (\ref{eq:89}) gives a proper distance to the singularity
that is roughly $\sqrt{0.9}$ times the distance in the metric
(\ref{eq:87}) at the same value of $-g_{00}$,
and the circumference of the spheres in the metric (\ref{eq:89})
is roughly $(1.2)^{1/6}$
the amount given by the metric (\ref{eq:87}),
but at least the qualitative behavior agrees.
Furthermore, the metric (\ref{eq:87}) is supposed to apply
at a distance $\ell$ that is less than one Planck length
from the naked singularity, and there one would expect
quantum gravity effects to change the form of the metric
or invalidate the use of a semiclassical metric altogether.
Therefore, I propose that the metric (\ref{eq:89})
may be taken as a reasonably good approximation to the
metric of an asymptotically flat static spherically
symmetric vacuum region when the backreaction of the
Boulware stress-tensor is self-consistently taken into account
in a semiclassical approximation,
and when one avoids the high curvature region where
the semiclassical approximation is expected to break down.

	We found in Eq. (\ref{eq:49})
that for a neutral spherical black hole
of area $A = 4\pi M^2$ surrounded by
a perfectly reflecting shell at $R-2M \ll M$,
the entropy is roughly
 \begin{equation}
 S \approx {1\over 4}A - {32\pi\alpha R\over R-2M}
   = {1\over 4}A - {n_b + {7 \over 8} n_f \over 360 (1-2M/R)}.
 \label{eq:90}
 \end{equation}
The last term represents the leading term
for the reduction of the entropy below one-quarter the area.
Let us ask how large this term can be
for various assumptions about the shell.

	First, consider the case that the shell
is held up entirely by its own stresses, with
no external forces (other than gravity) on it.
In particular, we shall consider the static shell
junction conditions (\ref{eq:37}) and (\ref{eq:38}),
applying the strong energy condition to the shell
so that its surface stress obeys the inequality
$S_2^2 = S_3^3 \leq -S_0^0$.  As we shall soon see,
it then turns out that $U_{-} = 1-2M_-/R \approx 1-W = 1-2M/R$
cannot be very small, so the terms involving the
pressures inside and outside the shell are then negligible.
Then the strong energy condition applied to the
junction conditions (\ref{eq:37}) and (\ref{eq:38})
imply that $U_{+}U_{-} \geq 1/25$, and since Eq. (\ref{eq:37})
implies that a shell with positive local mass has
$U_{+} < U_{-}$, we see that $1-W \approx U_{-} > 1/5$,
or $R > 2.5M$.  If Eq. (\ref{eq:90}) applied for such
a large value of $1-W$, it would then give
 \begin{equation}
 {1\over 4}A - S \approx  {n_b + {7 \over 8} n_f \over 360 (1-2M/R)}
 < {n_b + {7 \over 8} n_f \over 72},
 \label{eq:91}
 \end{equation}
a quite negligible decrease in the entropy,
unless somehow $n_b + (7/8) n_f$ is very large.

	Next, consider the case that the shell
has charge $Q$, so that its electrostatic repulsion holds it up.
Since we found above that the stresses within the surface
of the shell are quite ineffectual in holding up the shell
at $R-2M \ll M$, let us drop them from the junctions equations
but add the tension of the electromagnetic field
outside the shell and assume that that tension is much greater
than the radial pressures (or tensions) of the quantum fields.
Then the junction conditions (\ref{eq:37}) and (\ref{eq:38})
become
 \begin{equation}
 {\mu\over R} = U_{-}^{1/2} - U_{+}^{1/2}
 \label{eq:92}
 \end{equation}
and
 \begin{equation}
 0 = 8F = {\mu\over R} + U_{-}^{-1/2}
  - (1 - {Q^2\over R^2})U_{+}^{-1/2},
 \label{eq:93}
 \end{equation}.

	Now for fixed charge-to-mass ratio $Q/\mu$,
if we let $\gamma = (\mu/R)/U_{-}^{1/2} < 1$,
Eq. (\ref{eq:92}) implies that
$U_{+}^{1/2} = (1-\gamma)(\mu/R)$,
which when inserted into Eq. (\ref{eq:93}) gives
 \begin{equation}
 {1\over 1-2M/R} \approx {1\over U_{-}}
  = 1-\gamma+\gamma(Q/\mu)^2 < (Q/\mu)^2.
 \label{eq:94}
 \end{equation}
If we take the charge-to-mass ratio of an electron,
we get $(Q/\mu)^2 \approx 4.17\times 10^{42}$.
If we then suppose that somehow a shell of electrons
reflects electromagnetic (but not other) radiation
and thereby manages to keep the electromagnetic field
in its Boulware vacuum state outside the shell
(rather than in the Hartle-Hawking thermal state
that exists within the shell), then $n_b = 2$
(from the two helicities of photons) and $n_f=0$,
so one gets
 \begin{equation}
 {1\over 4}A - S \approx  {1 \over 180 (1-2M/R)}
 < 2.31\times 10^{40}.
 \label{eq:95}
 \end{equation}
 
 	Of course, there are severe problems in attaining
anything near this limit.  First, electrons in a shell
around a black hole, even if in static equilibrium
as I have calculated they can be, will not be in
stable equilibrium, and some unknown mechanism
would have to be invoked to keep the shell in place.
Second, without specifying how the electrons are to be kept in place,
it is hard to say how they will respond to the black hole thermal
radiation impinging upon them from below.
However, it is interesting that the upper limit
given by Eq. (\ref{eq:95}) for the reduction in the entropy
below one-quarter the (neutral) black-hole area,
from a shell held up by electrostatic forces,
is so large (because the charge-to-mass ratio
of an electron is so large).

	For a somewhat more nearly realistic
example of a shell around a black hole,
consider a thin aluminum foil that is charged
so that, like the shell of pure electrons,
the electrostatic forces balance the gravitational forces.
In this case there will be limitations from
the mass density $\rho$ of the foil,
the minimum practical thickness $\tau$ of the foil
and the maximum charge per surface area, $\sigma$,
that it can hold.
Again the possible tensile forces within the foil
itself are negligible in comparison with the
electrostatic forces and hence will be ignored.
(It turns out that the strong energy condition
also implies that they are not nearly sufficient
to stabilize the shell against radial perturbations,
which are unstable because the local gravitational forces
go up rapidly as the shell is brought closer to the
black hole horizon, whereas the electrostatic repulsion
forces depend only on the circumference of the shell,
which changes only slowly as the shell is moved
inward or outward near the horizon.
However, just as modern jet fighter planes
fly under the control of computer servomechanisms
while being deliberately constructed to be unstable,
and thus rapidly maneuverable,
I shall suppose that some unspecified servomechanism
can be used to keep the shell in place.
I shall leave to the reader the engineering problem
of constructing such a servomechanism
and just tell how to balance the forces in
the unstable equilibrium.)

	Let me now give some parameters associated with
the aluminum foil, using both conventional units, atomic units,
and Planck units (always the case in this paper when no units are
explicitly given).  For this discussion I shall use the charge
of the positron as
 \begin{equation}
 e \equiv e/\sqrt{4\pi\epsilon_0\hbar c} \approx 0.0854245329,
 \label{eq:96}
 \end{equation}
the mass of the positron or electron as
 \begin{equation}
 m \equiv m/\sqrt{\hbar c/G} \approx 4.185\times 10^{-23},
 \label{eq:97}
 \end{equation}
the mass of the proton as
 \begin{equation}
 m_p \equiv m_p/\sqrt{\hbar c/G} \approx 7.684\times 10^{-20},
 \label{eq:98}
 \end{equation}
the Rydberg energy as
 \begin{equation}
 E_R \equiv {1\over 2}m e^4 \approx 13.6057 \: eV
  \approx 1.114\times 10^{-27},
 \label{eq:99}
 \end{equation}
the Bohr radius as
 \begin{equation}
 a_B \equiv {1\over m e^2} \approx 5.2917721\times 10^{-9} {\rm cm}
  \approx 3.2755\times 10^{24},
 \label{eq:100}
 \end{equation}
the tropical year as
 \begin{equation}
 1 \ {\rm yr} \approx 5.854\times 10^{50},
 \label{eq:101}
 \end{equation}
and the solar mass as
 \begin{equation}
 M_{\odot} \approx 9.137\times 10^{37} \approx 0.5395/m_p^2.
 \label{eq:102}
 \end{equation}

	The density of solid aluminum is then
 \begin{equation}
 \rho \equiv N_{\rho}m_p/a_B^3 \approx 2.70 \:{\rm g}/{\rm cm}^2
  \approx 5.233\times 10^{-94},
 \label{eq:103}
 \end{equation}
with
 \begin{equation}
 N_{\rho} \approx 0.2391.
 \label{eq:104}
 \end{equation}
The aluminum will be so cold it will be superconducting,
in which case it has a Meissner magnetic penetration depth
of about 50 nm
\cite{Mei}.
As we shall see, the local Hawking temperature will be
far below the superconducting gap energy, so the foil
will be almost completely reflecting to the thermal
electromagnetic radiation if its thickness is several
times the magnetic penetration depth.
To be very conservative, I shall take the thickness
of the foil to be about 100 times the magnetic penetration depth,
 \begin{equation}
 \tau \equiv N_{\tau}a_B
  = 50 \:{\rm microns} = 0.0005 \:{\rm cm} \approx 3.094\times 10^{29},
 \label{eq:105}
 \end{equation}
with
 \begin{equation}
 N_{\tau} \approx 94\,486.
 \label{eq:106}
 \end{equation}

	I shall choose the electric surface charge density
$\sigma$ so that if it were an excess of electrons,
the probability for one to tunnel off, say $P$,
is very small, say $\exp{(-100)}$ in some suitable atomic
time unit.  Since $P \sim \exp{(-2I)}$ with tunneling amplitude
$I$, I shall choose $I = (1/2)\ln{(1/P)} = 50$.
Now the work function for polycrystalline aluminum is
 \begin{equation}
 V_0 \equiv v_0 E_R \approx 4.28 \: eV,
 \label{eq:107}
 \end{equation}
with
 \begin{equation}
 v_0 \approx 0.315.
 \label{eq:108}
 \end{equation}
If one takes the potential energy of an electron,
relative to that at the Fermi surface, to be
$V_0 - e E x$ at distance $x$ from the surface,
where $E = 4\pi\sigma$ is the magnitude of the charge
density (dropping the minus sign that $\sigma$
would have in actuality, because of Ben Franklin's
inconvenient sign convention, if there were an excess
of electrons on the surface),
then the WKB amplitude for the electron to tunnel
through the classically forbidden region
$0 < x < x_0 = V_0/(eE)$ is
 \begin{equation}
 I = \int_0^{x_0}\sqrt{2mV}dx = {\sqrt{8mV_0^3}\over 3eE}.
 \label{eq:109}
 \end{equation}
This then gives
 \begin{eqnarray}
 &\sigma& \equiv {m^2 e^5\over N_{\sigma}}
  = {m^2 e^5 v_0^{3/2}\over 6\pi\ln{(1/P)}}
 \approx {m^2 e^5 v_0^{3/2}\over 600\pi}
 \nonumber \\
 &\approx& 3.34\times 10^{12} \: e/{\rm cm}^2
 \approx 0.00893 \: e/a_B^3
 \approx 7.46\times 10^{-55},
 \label{eq:110}
 \end{eqnarray}
with
 \begin{equation}
 N_{\sigma} = 6\pi\ln{(1/P)}v_0^{-3/2}
 = 600\pi v_0^{-3/2} \approx 10\,700.
 \label{eq:111}
 \end{equation}
Expressed in terms of the unit area $n^{-2/3}$
formed from the aluminum atomic number density $n$,
one needs about one excess electron per 460 of
these unit areas to give this surface charge density
$\sigma$, so this does not seem excessive.

	From these parameters, one gets
that the charge-to-mass ratio of the aluminum foil is
 \begin{equation}
 {Q\over\mu}={\sigma\over\rho\tau}
 ={e/m_p\over N_{\sigma}N_{\rho}N_{\tau}}
 \approx {1.112\times 10^{18}\over 2.41\times 10^8}
 \approx 4.61\times 10^9.
 \label{eq:112}
 \end{equation}
This is that of the pure electron shell, divided by
$N_{\sigma}N_{\rho}N_{\tau}(m_p/m) \approx 4.43\times 10^{11}$.

	Then, by the same analysis used above for
the pure electron shell, one finds that if one takes
$\mu/R = U_{-}^{1/2} = \mu/Q$, one gets
 \begin{equation}
 {1\over 1-2M/R} = \left({Q\over\mu}\right)^2
 \approx 2.12\times 10^{19},
 \label{eq:113}
 \end{equation}
and hence the reduction of the entropy from excluding
the thermal photons from above the shell is
 \begin{eqnarray}
 \Delta S \equiv {1\over 4}A - S &\approx& {1 \over 180 (1-2M/R)}
 \approx {(e/m_p)^2\over 180 (N_{\sigma}N_{\rho}N_{\tau})^2}
 \nonumber \\
 &\approx& {1.24\times 10^{36}\over 1.05\times 10^{19}}
 \approx 1.18\times 10^{17}.
 \label{eq:114}
 \end{eqnarray}
As we shall see, this is small in comparison with
the total entropy of the black hole, but the reduction
means that the number of states is fewer by a factor
of about
 \begin{equation}
 e^{\Delta S} \sim 10^{51\,000\,000\,000\,000\,000},
 \label{eq:115}
 \end{equation}
which is quite a large factor.

	One can further derive properties of the black hole
around which this aluminum foil is placed to minimize
$1-2M/R$ and hence maximize $\Delta S$.
The radius of the shell (which is very nearly that of the hole) is
 \begin{equation}
 R = {1\over 4\pi\sigma} \approx 1.07\times 10^{53}
  \approx 182 \ {\rm light \ years},
 \label{eq:116}
 \end{equation}
giving a circumference of about 1150 light years.
If the redshift factor to infinity were $\sqrt{1-2M/R}$,
the time as seen at infinity for a photon to circumnavigate the shell
would be about $5.3 \times 10^{12}$ years, hundreds of times
longer than the current age of the universe.
(In actuality, if one does take the limit of setting
$\mu/R = U_{-}^{1/2} = \mu/Q$, one finds that as seen from
the outside, the shell is at what would be the horizon
of an extreme Reissner-Nordstrom black hole,
so there would be an infinite redshift factor to infinity.
But if one set $\mu/R$ to be, say, half as large,
the previous quantities would be shifted by merely
factors of two or so, and then the time as seen at infinity
for a photon to circumnavigate the foil shell would be
of the order of hundreds of times the present age of the universe.)

	The mass of the black hole is then
 \begin{equation}
 M \approx {1\over 2}R = {1\over 4\pi\sigma} \approx 5.34\times 10^{52}
  \approx 5.84\times 10^{14} M_{\odot},
 \label{eq:117}
 \end{equation}
of the order of mass of a supercluster of galaxies.
This mass then gives a total entropy for the black hole of
 \begin{equation}
 S \approx 4\pi M^2 \approx 3.58\times 10^{106},
 \label{eq:118}
 \end{equation}
which is about $3.04 \times 10^{89}$ times the reduction $\Delta S$
in the entropy calculated above.  Therefore, as already mentioned,
the relative reduction of the entropy is negligible
in this case, but it does reduce the total number of states
(far, far more than a googolplex in this example)
by the huge factor given by Eq. (\ref{eq:115}).

	The proper radial distance of the shell to the horizon is
 \begin{equation}
 D \approx 2M\sqrt{1-2M/R} \approx 2.50 \ {\rm light \ seconds}.
 \label{eq:119}
 \end{equation}
The local acceleration of gravity as seen from just inside the shell is
 \begin{equation}
 g \approx {c^2\over D} \approx 1.20\times 10^8 \: {\rm m}/{\rm s}^2
 \approx 1.22\times 10^7 \: g_{\oplus},
 \label{eq:120}
 \end{equation}
over twelve million times the acceleration of gravity $g_{\oplus}$
at the surface of the earth.

	The local Hawking temperature as seen at the inner edge
of the shell is
 \begin{equation}
 T_{\rm local} \approx {1\over 2\pi D} \approx 3.44\times 10^{-45}
  \approx 4.87\times 10^{-13} \: K,
 \label{eq:121}
 \end{equation}
which is indeed far below the critical superconducting temperature
of 1.175 K
\cite{Rob},
so the shell should stay superconducting and indeed reflect
almost all of the electromagnetic radiation emitted by the black hole.

	Thus we have seen that by placing a reflecting shell
around a black hole, we can make the entropy have a value
that is below one-quarter its area.
If we are allowed an idealized perfectly reflecting shell
that can be placed within roughly one Planck length of the horizon,
then this entropy reduction can be of the same order as
the area of the hole.  For a more realistic shell,
such as a superconducting aluminum foil,
the entropy reduction can only be a tiny fraction of the area,
but it still can be huge in absolute units,
markedly reducing the number of black hole states from
what would be erroneously estimated by exponentiating
one quarter the horizon area.

	I have benefited from conversations with,
among many others whose names did not immediately come to mind,
Valeri Frolov, Frank Hegmann, Akio Hosoya, Satoshi Iso, Sang-Pyo Kim,
Frank Marsiglio, Sharon Morsink, Shinji Mukohyama, Jonathan Oppenheim,
L. Sriramkumar,and Andrei Zelnikov.
Part of this work was done at the Tokyo Institute
of Technology under the hospitality of Akio Hosoya,
and part was done at the Haiti Children's Home
of Mirebalais, Haiti, under the hospitality
of Patricia and Melinda Smith while adopting six-year-old
Marie Patricia Grace to become our fifth child.
This research was supported in part by
the Natural Sciences and Engineering Research
Council of Canada.

\end{document}